\documentclass{bmvc2k}

\title{A self-supervised and adversarial approach to hyperspectral demosaicking and {RGB} reconstruction in surgical imaging}

\addauthor{Peichao Li}{peichao.li@kcl.ac.uk}{1}
\addauthor{Oscar MacCormac}{oscar.j.maccormac@kcl.ac.uk}{1,2}
\addauthor{Jonathan Shapey}{jonathan.shapey@kcl.ac.uk}{1,2}
\addauthor{Tom Vercauteren}{tom.vercauteren@kcl.ac.uk}{1}

\addinstitution{
 School of Biomedical Engineering \& Imaging Sciences\\
 King's College London\\
 London, UK
}
\addinstitution{
 Department of Neurosurgery\\
 King's College Hospital NHS Foundation Trust,\\
 London, UK
}

\runninghead{Li et al.}{Self-supervised demosaicking and RGB reconstruction}

\usepackage{amssymb}
\usepackage{pdfpages}
\usepackage{enumitem}
\setlist{noitemsep,topsep=3pt,before=\vspace{-\topsep}}

\begin{document}

\maketitle

\begin{abstract}
Hyperspectral imaging holds promises in surgical imaging by offering biological tissue differentiation capabilities with detailed information that is invisible to the naked eye.
For intra-operative guidance, real-time spectral data capture and display is mandated.
Snapshot mosaic hyperspectral cameras are currently seen as the most suitable technology given this requirement.
However, snapshot mosaic imaging requires a demosaicking algorithm to fully restore the spatial and spectral details in the images.
Modern demosaicking approaches typically rely on synthetic datasets to develop supervised learning methods, as it is practically impossible to simultaneously capture both snapshot and high-resolution spectral images of the exact same surgical scene.
In this work, we present a self-supervised demosaicking and RGB reconstruction method that does not depend on paired high-resolution data as ground truth.
We leverage unpaired standard high-resolution surgical microscopy images, which only provide RGB data but can be collected during routine surgeries.
Adversarial learning complemented by self-supervised approaches are used to drive our hyperspectral-based RGB reconstruction into resembling surgical microscopy images and increasing the spatial resolution of our demosaicking.
The spatial and spectral fidelity of the reconstructed hyperspectral images have been evaluated quantitatively. Moreover, a user study was conducted to evaluate the RGB visualisation generated from these spectral images.
Both spatial detail and colour accuracy were assessed by neurosurgical experts.
Our proposed self-supervised demosaicking method demonstrates improved results compared to existing methods, demonstrating its potential for seamless integration into intra-operative workflows.
\end{abstract}

\section{Introduction}
\label{sec:intro}
Hyperspectral imaging (HSI) is becoming a widely used technique which collects information across a broad range of the electromagnetic spectrum. It perceives colours beyond human vision and can provide valuable medical data, including tissue perfusion, oxygen saturation, and other diagnostic measurements \cite{holmer2018hyperspectral}. As a non-contact, non-ionizing, and non-invasive method, it offers advantages for numerous medical applications \cite{lu2014medical,shapey2019intraoperative,clancy2020surgical} and various groups have investigated how best to integrate HSI systems into surgical workflows \cite{fabelo2018intraoperative,shapey2019intraoperative,fabelo2019helicoid,ebner2021intraoperative,maccormac2023lightfield}. For consistency, in this work, we will use the term hyperspectral imaging, although it sometimes called multispectral imaging depending on the number of bands.

Similar to Bayer Colour Filter Arrays (CFA) for RGB imaging, snapshot mosaic HSI uses Multi-spectral Filter Arrays (MSFA) to capture multiple spectral bands in a single exposure with each pixel being exposed to a single fixed spectral band.
Its rapid acquisition time and compact camera system enables seamless integration into neuro-oncology surgical environments \cite{li2022deep}.
Typically, MSFA sensors are arranged in $4 \times 4$ or $5 \times 5$ mosaic patterns to simultaneously obtain 16 or 25 spectral channels respectively.
Snapshot mosasic HSI systems thus sacrifices spatial and spectral resolution to achieve real-time spectral image acquisition.
Consequently, a high-quality demosaicking step, more critical than standard debayering for RGB CFAs, is required to recover the lost spatial information
Details on snapshot mosaic imaging and demosaicking can be found in our previous work in \cite{li2022deep}.

Traditional spectral demosaicking methods typically rely on interpolation-based techniques \cite{yu2006colour,mizutani2014multispectral,mihoubi2017multispectral, ogawa2016demosaicking,gupta2022adaptive} or model optimisation approaches \cite{eismann2004application,tsagkatakis2019graph,xue2019enhanced,xue2021multilayer}.
However, these methods generally under-perform compared to learning-based methods.
Recent deep learning methods have demonstrated their effectiveness in image super-resolution tasks \cite{dong2014learning,lim2017edsr,zhang2018image,lugmayr2020ntire,zhang2023ntire}. Consequently, similar approaches are now being investigated for application to the hyperspectral image demosaicking problem. Existing learning-based methods depend on synthetic snapshot images paired with ideal reconstructed images to create datasets for supervised training \cite{mei2017hyperspectral,habtegebrial2019deep,dijkstra2019hyperspectral, feng2021mosaic,arad2022ntire,li2022deep}. However, these synthetic hyperspectral images often fail to capture the complexities of real-world scenarios due to simplified or idealised conditions during generation \cite{dijkstra2019hyperspectral,feng2024unsupervised}. Therefore, these methods may generalise poorly when applied to actual datasets.

In our recent work~\cite{li2023spatial}, we adopted an alternative approach, posing demosaicking as an ill-posed inverse problem, and developed a self-supervised learning-based hyperspectral demosaicking algorithm that relies solely on snapshot mosaic data and the physics of acquisition. In this study, we introduced a Spatial Gradient Consistency (SGC) term as a loss function for self-supervised training, which promotes cross-band correlation and enhances image detail. 
However, spectral demosaicking networks are often prone to periodic gridding artifacts due to over-fitting~\cite{feng2024unsupervised}.
\citet{li2023spatial} attempted to mitigate this issue by incorporating Tikhonov regularisation and total variation into the loss function.
This approach nonetheless introduced a trade-off between image sharpness and smoothing. 

Generative Adversarial Networks (GANs)~\cite{goodfellow2014generative} might be another promising solution for effectively alleviating repetitive artifacts while preserving spatial details.
GANs can be trained with unpaired data and have demonstrated efficacy in single-image super-resolution \cite{ledig2017photo, sajjadi2017enhancenet, wang2018esrgan} and debayering tasks~\cite{dong2018joint, luo2020image, qu2022demosaicing}.
GAN-based methods typically employ generator networks that map low-resolution source images to high-resolution results, along with discriminators that assess both generated results and instances from the target distribution to determine their authenticity.
The application of GAN-based approaches to hyperspectral image demosaicking, especially in surgical hyperspectral imaging, remains limited.
Indeed, even with unpaired data allowance, acquiring
a large number of high-resolution hyperspectral images in a surgical environment is challenging~\cite{fabelo2019helicoid}.

In contrast to spectral data, obtaining high-resolution RGB images intra-operatively with neurosurgical microscopes is relatively easier.
In this work, we propose to exploit RGB images from the surgical microscopes to aid in the demosaicking of surgical hyperspectral images using GANs.
The conversion of hyperspectral images to RGB is a physically explainable process, which will be discussed in Section \ref{sec:conversion}. Additionally, convincing spectral reconstruction from RGB images has been demonstrated \cite{koundinya20182d3d,li2020adaptive,arad2022recovery}
and these can be used in a GAN-based approach exploiting cycle-consistency criteria~\cite{zhu2017unpaired,yuan2018unsupervised,ravi2019adversarial}.
These existing research makes it possible to develop a GAN-based method to assist hyperspectral demosaicking with the help of RGB images.

This work proposes a real-time neurosurgical hyperspectral image demosaicking algorithm that leverages unpaired surgical microscopy RGB images to enhance the quality of the reconstructed hyperspectral images. Our contributions are threefold:
\begin{itemize}
    \item We introduce a cycle-consistent adversarial loss to use high-resolution RGB images for enhancing details and mitigating potential artifacts in the reconstructed images.
    \item We propose replacing the physics-based hyperspectral-to-RGB conversion operation of \cite{li2023spatial} with a simplified neural network model to further improve the colour accuracy of the RGB visualisations of the hyperspectral images, thereby more closely resembling surgical microscopy images to aid surgeons in decision-making.
    \item We introduce an inverse pixel shuffle loss term designed to eliminate periodic gridding artifacts more effectively than Tikhonov regularisation, as previously used in \cite{li2023spatial}, while preserving local spatial details.
\end{itemize}

The algorithm is developed using actual neurosurgical snapshot mosaic images and high-resolution surgical microscopy images, and evaluated both quantitatively and qualitatively. These evaluations validate its potential to be integrated into real-time surgical hyperspectral systems, assisting clinical workflows and transitioning into regular clinical practice.

\section{Methodology}
\begin{figure*}
\begin{center}
\fbox{\includegraphics[width=0.9\textwidth]{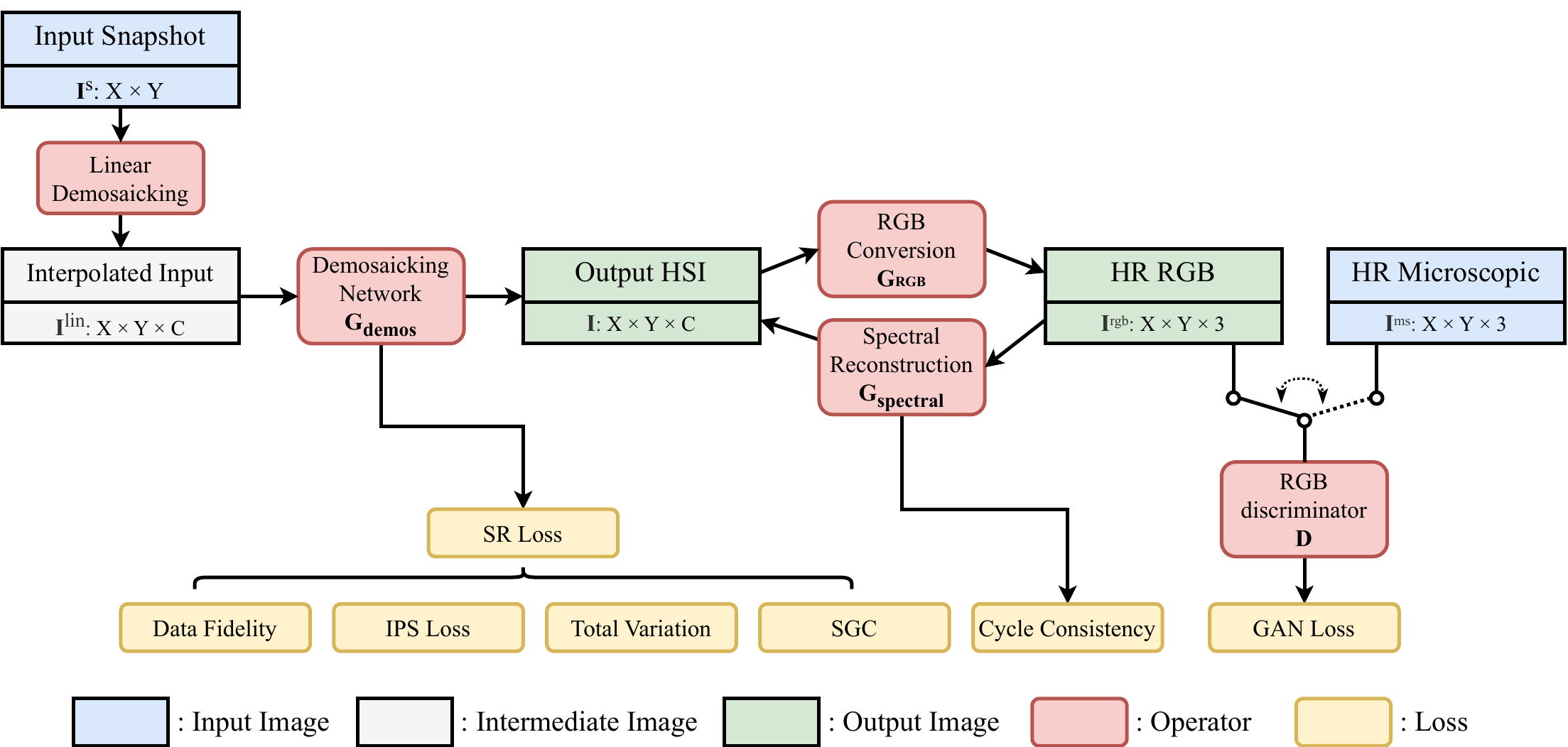}}
\end{center}
   \caption{Diagram of the proposed demosaicking algorithm. First, bilinear interpolation is applied to the input snapshot mosaic image to recover a fully-sampled spatial and spectral grid. The interpolated image then serves as input to the demosaicking network $G_{\text{demos}}$, which generates the refined hyperspectral image. Super-resolution losses and adversarial losses are computed for training the network.}
\label{fig:diagram-demosaicking}
\end{figure*}

The overall framework of our proposed demosaicking algorithm is summarised in Figure~\ref{fig:diagram-demosaicking}. First, bilinear interpolation is applied to the input snapshot mosaic image to recover a fully-sampled spatial and spectral grid.
The interpolated image then serves as input to the demosaicking network $G_{\text{demos}}$, which generates the refined hyperspectral image.
Many deep neural networks suitable for image super-resolution or demosaicking can be adapted for \( G_{\text{demos}} \).
In our work, the modified Res2-Unet model \cite{song2022hyperspectral} was chosen
as it demonstrated outstanding performance in the NTIRE 2022 demosaicking challenge \cite{arad2022ntire} and has proven effective when adapted to self-supervised hyperspectral demosaicking tasks~\cite{li2023spatial}.
Similar to \cite{li2023spatial}, an overriding operator is embedded in the demosaicking network to ensure that pixels from the original snapshot image are directly incorporated into the output image, thus maintaining perfect data fidelity.

\subsection{Cycle-consistent adversarial training}
\label{sec:gan}

Our demosaicking model \( G_{\text{demos}} \) along with the RGB conversion model \( G_{\text{RGB}} \) aim to generate images that look similar to a surgical microscopy image.
Similar to CycleGAN~\cite{zhu2017unpaired}, the combination of $G_{\text{demos}}$ and $G_{\text{RGB}}$ is used as a generator network aiming at deceiving an RGB discriminator \( D \) trained to distinguish between real and generated RGB images.
A $70 \times 70$ PatchGAN discriminator~\cite{isola2017image} was used for \( D \) as per CycleGAN~\cite{zhu2017unpaired} and CinCGAN~\cite{yuan2018unsupervised}.
To stabilise the training procedure, we replace the negative log-likelihood with the least squares in the adversarial loss as suggested by \cite{mao2016multi}:
\begin{equation}
    \mathcal{L}_{\text{GAN}} = \mathbb{E}[\| D(I^{\text{RGB}}) - 1 \|_2] + \mathbb{E}[\| D(G_{\text{RGB}}(G_{\text{demos}}(I^{\text{lin}}))) \|_2]
\end{equation}
For simplicity, we do not detail the adversarial training procedure but, as typical, a min-max approach is used to iteratively train the generator and discriminator. 

To ensure consistency between the RGB reconstruction and the original hyperspectral image, a spectral recovery model \( G_{\text{spectral}} \) is used to convert the RGB image back to the hyperspectral image, and a cycle consistency loss is introduced:
\begin{equation}
    \mathcal{L}_{\text{cyc}} = \mathbb{E}[\| G_{\text{spectral}} (G_{\text{RGB}} (G_{\text{demos}} (I^{\text{lin}}))) - G_{\text{demos}} (I^{\text{lin}}) \|_1]
\end{equation}

\subsection{Converting Between Hyperspectral and RGB Images}
\label{sec:conversion}
\begin{figure}
\begin{tabular}{ccc}
\bmvaHangBox{\fbox{\includegraphics[width=2.6cm]{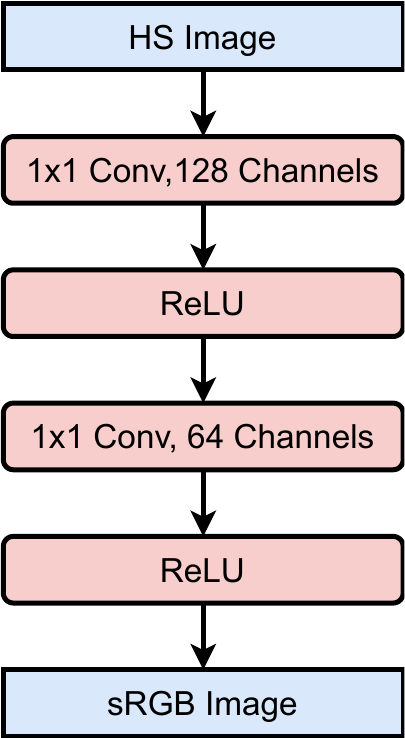}}}&
\bmvaHangBox{\fbox{\includegraphics[width=4.75cm]{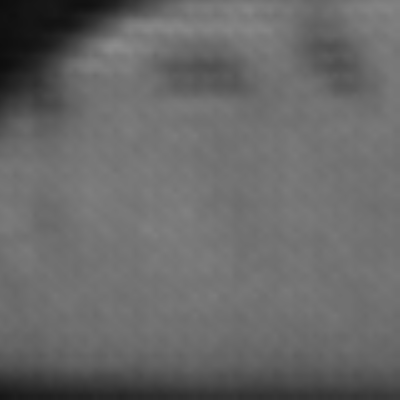}}}&
\bmvaHangBox{\fbox{\includegraphics[width=3.2cm]{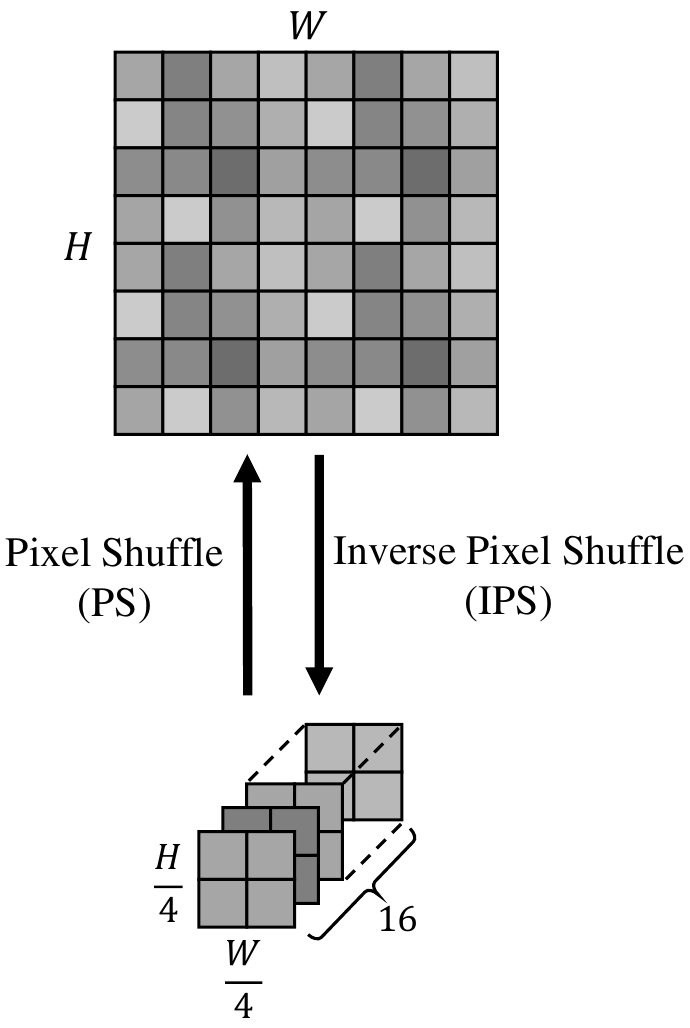}}}\\
(a)&(b)&(c)
\end{tabular}
\caption{(a) MLP model for hyperspectral-to-RGB conversion.
(b) An example of periodic gridding artefacts that can be observed with state-of-the-art hyperspectral demosaicking \cite{li2023spatial}.
(c)~Diagram illustrating the Inverse Pixel Shuffle (IPS) operation.}
\label{fig:methods}
\end{figure}

Traditionally, converting hyperspectral images to RGB images involves a colour matching function that maps spectral data to the CIE 1931 XYZ colour space \cite{foster2019hyperspectral,garcia2023hyperspectral}.
A linear transformation then converts CIE XYZ images to linear standard RGB (sRGB) and gamma correction is applied to produce the sRGB image. Further details on this process can be found in our previous paper~\cite{li2022deep}. 
However, the colour of the RGB visualisation of the snapshot hyperspectral image may not accurately represent actual appearances, particularly in the red spectrum, which is prevalent in many neurosurgical images. This limitation of sRGB reconstruction may be related to the relatively coarse spectral resolution and relatively narrow spectral coverage of snapshot mosaic sensors. It may also be explained by proprietary colour corrections not captured in the sRGB convention but embedded in commercial surgical microscope cameras.

Here, we replace the traditional fixed sRGB operations with a simple trainable neural network model $G_{\text{RGB}}$.
This model uses a multilayer perceptron (MLP), illustrated in Figure \ref{fig:methods}(a), that expands the spectral signature at each pixel using $1 \times 1$ convolutions, then reduces them to 3 channels to derive the RGB values. The use of $1 \times 1$ convolutions ensures that the model focuses solely on RGB conversion without modifying the spatial details.

The conversion from RGB image back to hyperspectral image remains a complex challenge, but numerous research exists on this topic. In our work, we have adopted the Adaptive Weighted Attention Network (AWAN) \cite{li2020adaptive} for \( G_{\text{spectral}} \). AWAN adaptively recalibrates channel-wise feature responses and effectively captures correlations in distant regions, thereby enhancing the spectral recovery process.

\subsection{Inverse Pixel Shuffle loss}
\label{sec:ips}
Hyperspectral demosaicking networks are prone to periodic gridding artifacts as shown in Figure \ref{fig:methods}(b). \citet{feng2024unsupervised} proposed to apply the Inverse Pixel Shuffle (IPS) method \cite{shi2016real} to reorganise demosaicked band images into several sub-images as shown in Figure \ref{fig:methods}(c). They compute the variance in the global mean of all sub-images, and use it as a post-training metric to quantify over-fitting artifacts and select the best-fitting model. Building on their approach, we turn the IPS metric into a back-propagatable loss function used during training to minimise gridding artifact:
\begin{equation}
    \mathcal{L}_{\text{IPS}} = \frac{1}{B} \sum_{b=1}^{B} \operatorname{Var}_{\textrm{channels}}(\operatorname{Mean}_{\textrm{space}}(\text{IPS}(I_b)))
\end{equation}
where \( B \) represents the number of spectral bands and \( I_b \) denotes the reconstructed single-channel image corresponding to the \( b \)-th spectral band.
This loss ensures that the global statistics of all sub-images after IPS remain consistent, yet permits local variances that enhance the spatial details of the image, thus effectively alleviating periodic gridding artifacts. 

Finally, we combine the above losses with additional regularisation from Total Variation and Spatial Gradient Consistency (SGC) \cite{li2023spatial}. 
The total loss function used to train the networks is defined as follows:
\begin{equation}
    \mathcal{L}_{\text{total}} = \mathcal{L}_{\text{SGC}} + \lambda_{\text{TV}} \mathcal{L}_{\text{TV}} + \lambda_{\text{IPS}} \mathcal{L}_{\text{IPS}} + \lambda_{\text{GAN}} \mathcal{L}_{\text{GAN}} + \lambda_{\text{cyc}} \mathcal{L}_{\text{cyc}}
    \label{eq:total-loss}
\end{equation}
where \(\lambda_{\text{TV}}\), \(\lambda_{\text{IPS}}\), \(\lambda_{\text{GAN}}\), and \(\lambda_{\text{cyc}}\) are the weighting coefficients for the Total Variation (TV) loss, Inverse Pixel Shuffle (IPS) loss, adversarial (GAN) loss, and cycle consistency loss, respectively.
As mentioned above, no data fidelity loss is needed as perfect data fidelity is ensured by an overriding operator embedded in the demosaicking network. Also, a min-max optimisation is used to alternatively train the discriminator and the other networks.

\section{Experiments and Results}
\subsection{Surgical imaging data}
The data was obtained from patients undergoing neurosurgery as part of a single-centre prospective clinical observational study involving a prototype hyperspectral imaging system (NeuroHSI study: REC reference 22/LO/0046, ClinicalTrials.gov ID NCT05294185).
All patients provided informed consent. This study assesses the intra-operative capabilities of a $4 \times 4$, 16-band visible range snapshot mosaic camera (IMEC CMV2K-SSM4X4-VIS) to characterise neurosurgical tissue.

A total of 210 snapshot mosaic image frames, showing minor motion blur or out-of-focus blur, were manually selected from the video data. These frames were taken from 17 different surgical cases, including both neuro-oncology and neuro-vascular procedures. Of these, 150 images from 8 cases were manually selected for training, 30 images from 4 cases for validation, and the remaining 30 images from 5 cases for testing. While no high-resolution hyperspectral images were acquired for ground truths, high-resolution RGB images were obtained using a ZEISS KINEVO 900 neurosurgical microscopes. From these, 210 surgical microscopy images were also manually selected. To challenge the discriminator, these images were intentionally chosen to depict similar scenes with similar anatomical structures or surgical tools presented as the corresponding hyperspectral images, even though they were not aligned.

\subsection{Training details}
As GAN training can be very unstable, all networks require careful initialisation. The training snapshot data are initially demosaicked linearly and then converted to RGB image using the traditional method described in Section~\ref{sec:conversion}, which uses CIE XYZ as the intermediate step.
This process generates a dataset of matched hyperspectral and RGB image pairs, which we use to pre-train the hyperspectral-to-RGB MLP model \( G_{\text{RGB}} \) and the AWAN model \( G_{\text{spectral}} \).
The MLP model was trained using the L1 loss with an Adam optimiser and an initial learning rate of \(1 \times 10^{-2}\), while the AWAN model pre-training followed the training protocol outlined in~\cite{li2020adaptive}.
We initialised the Res2-Unet demosaicking model using the trained weights from our previous SGC-based self-supervised method~\cite{li2023spatial}.
In our experiments, we observed that the discriminator model \( D \) learned very rapidly, so we simply applied Xavier initialisation to this model without using pre-trained parameters.

The pre-trained models together with the Xavier-initialised discriminator served as a starting point for subsequent fine-tuning where all models are trained jointly.
The weighting coefficients, as outlined in Equation~\eqref{eq:total-loss}, were set as follows: $\lambda_{\text{TV}} = 1 \times 10^{-3}$, $\lambda_{\text{IPS}} = 1$, $\lambda_{\text{SGC}} = 1$, $\lambda_{\text{GAN}} = 0.1$, and $\lambda_{\text{cyc}} = 1$. The Adam optimiser was retained for fine-tuning.
The initial learning rates were adjusted to $1 \times 10^{-5}$ for the Res2-Unet, $1 \times 10^{-3}$ for the MLP, and $1 \times 10^{-6}$ for the AWAN. All algorithms were implemented using PyTorch.

\begin{table}
\begin{center}
\begin{tabular}{|l|c|c|}
\hline
Method & BRISQUE & FID Score \\ 
\hline
Linear & $67.12 \pm 5.04$ & $104.57$ \\
PPID & $59.65 \pm 5.03$ & $98.95$ \\
GRMR & $40.55 \pm 7.73$ & $103.20$ \\
SGC & $31.56 \pm 9.35$ & $85.29$ \\
Ours (w.o. RGB model) & $19.73 \pm 5.99$ & $78.93$ \\
Ours (with RGB model) & $19.35 \pm 5.77$ & $75.62$ \\
\hline
\end{tabular}
\end{center}
\caption{Average quantitative results measuring the quality of the RGB visualisation of the demosaicked hyperspectral images.
Lower is better for both BRISQUE and FID.
\label{tab:quantitative}}
\end{table}

\subsection{Quantitative results}
\begin{figure*}[tb!]
\begin{center}
\fbox{\includegraphics[width=0.9\textwidth]{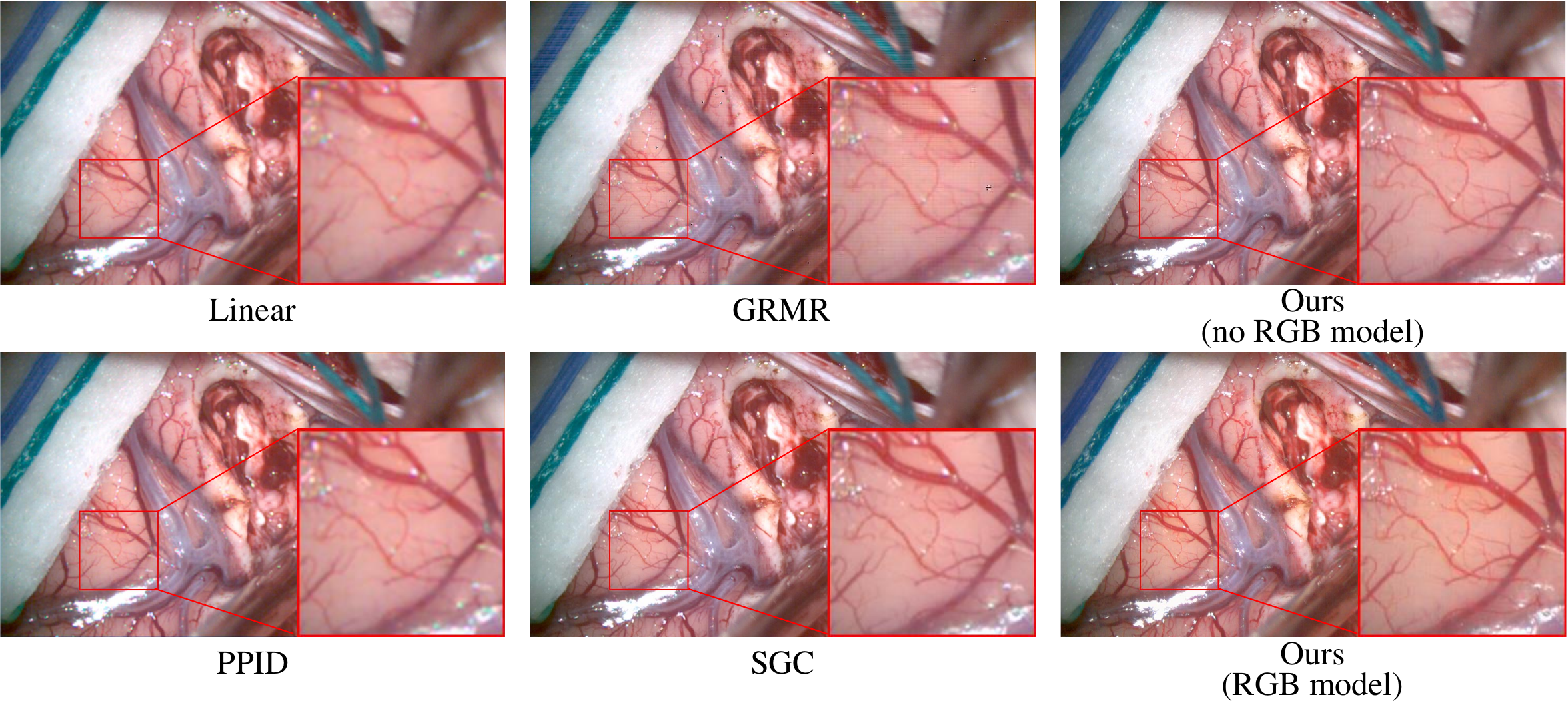}}
\end{center}
   \caption{Comparison between different demosaicking methods on an example NeuroHSI test image.}
\label{fig:all-results}
\end{figure*}

We compared our proposed demosaicking methods with three existing methods that do not rely on high-resolution hyperspectral data for ground truths, including PPID \cite{mihoubi2017multispectral}, GRMR~\cite{tsagkatakis2019graph}, and SGC \cite{li2023spatial}.
PPID is an interpolation-based demosaicking method based on the pseudo-panchromatic image.
GRMR is an iterative optimisation approach using a low-rank and graph regularised optimisation framework. Both PPID and GRMR are computational methods that do not require training. SGC is our previous state-of-the-art self-supervised demosaicking method, which employs the novel SGC term along with traditional regularisation terms. To evaluate the SGC method and compare it with our proposed algorithm, we trained a Res2-Unet model following the protocol outlined in the original paper on the same dataset as our proposed algorithm in this experiment.
Irrespective of whether the algorithm was trained jointly with an RGB model, to enable a fair comparison of the spatial reconstruction quality only, 
all hyperspectral demosaicking results were converted to RGB images using the traditional method outlined in Section~\ref{sec:conversion}.
For additional colour fidelity analysis, 
we compare the result of our full model that includes the trainable RGB conversion network to the one used for spatial reconstruction quality where the RGB reconstruction is performed with a standard fixed sRGB approach.
Figure~\ref{fig:all-results} shows the comparison between different demosaicking methods on an example test image. Given the absence of ground truth data for comparison, we adopted the non-reference image quality evaluation metric, BRISQUE \cite{mittal2012no}, to assess the quality of the outputs.
Furthermore, as unpaired high-resolution surgical microscopy images were available as reference data, we also computed the Fréchet Inception Distance (FID) score \cite{heusel2017gans} to evaluate the similarity between the surgical microscopy images and the RGB representations of the demosaicked results.

The average BRISQUE and FID scores for different methods are presented in Table \ref{tab:quantitative}. Our proposed demosaicking algorithm significantly outperforms the other methods in terms of BRISQUE score, achieving a two-tailed p-value of less than \(10^{-7}\) when conducting a t-test against all other methods. 
Additionally, our algorithm achieved the lowest FID score, indicating that the RGB visualisations of our demosaicked results are more closely aligned with surgical microscopy images compared to those from other methods. We also perform an ablation study to show the effectiveness of the different loss terms in \eqref{eq:total-loss}. The details can be found in the supplementary materials.

\begin{figure*}[tb!]
\begin{center}
\fbox{\includegraphics[width=0.9\textwidth]{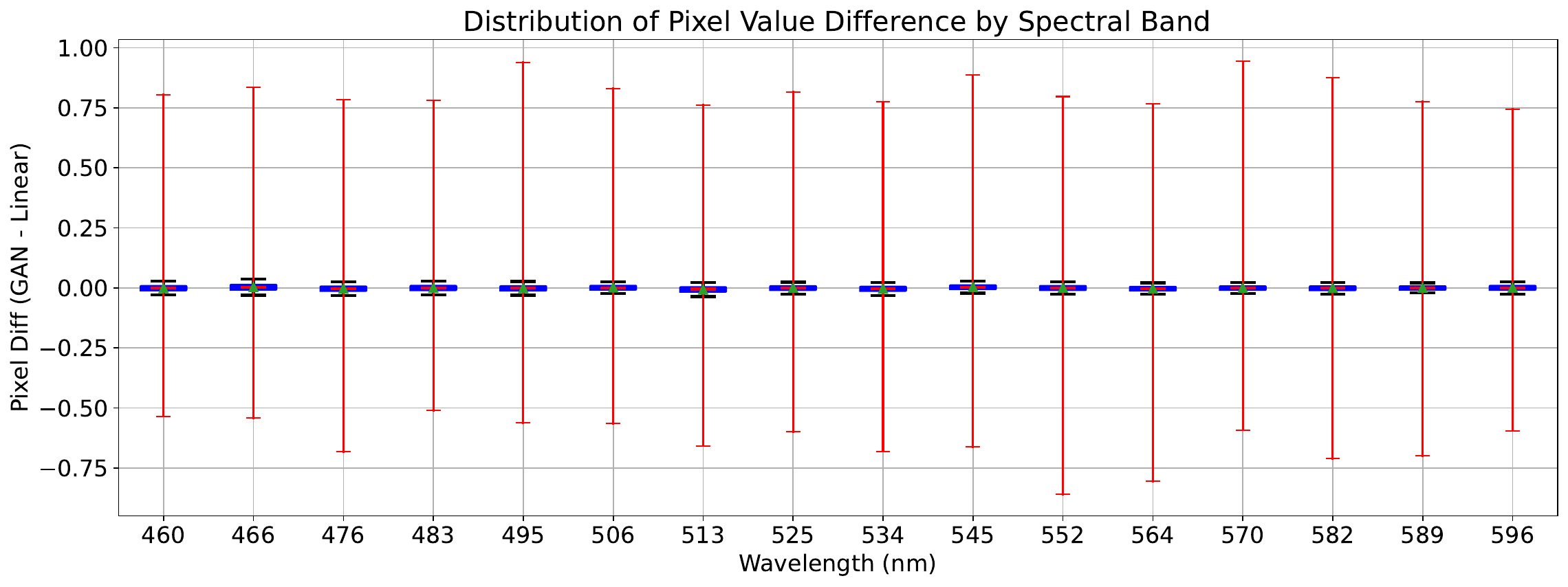}}
\end{center}
   \caption{Box plot illustrating the difference between our demosaicking results and linear demosaicking.
   Differences are expected whenever edges occur in the images as linear demosaicking otherwise results in low-resolution reconstruction. This analysis shows the absence of spectral bias in our reconstructions.}
\label{fig:box-plot}
\end{figure*}

To assess the spectral accuracy and absence of bias in the hyperspectral images demosaicked with our proposed algorithm, we analysed the pixel value differences between our results and the linear demosaicking results.
While not a ground truth, linear demosaicking can indeed be considered as an unbiased spectral reconstruction approach despite its poor ability to reconstruct spatial details. 
These differences are depicted in a box plot, as shown in Figure~\ref{fig:box-plot}, where most differences cluster around zero, indicating minimal deviation. Notably, only a few pixels exhibit significant discrepancies from linear demosaicking, as shown in the red lines in Figure~\ref{fig:box-plot} as "outliers", which is to be expected as image edges are poorly reconstructed with linear interpolation based demosaicking.
A plot illustrating the per-band pixel differences for one of the test images are presented in the supplementary material, where it reveals that larger differences tend to occur only in regions where our algorithm enhances spatial details, thereby confirming that these are not indicative of any spectral shift.

In our experiment, processing a single snapshot image with dimensions $1280\times 720$ takes 41 ms on average.
This measurement was obtained by demosaicking 100 images using the Res2-Unet for generating hyperspectral images and our proposed RGB model for RGB visualisation, on an NVIDIA RTX 4070 Ti Super. These results highlight that our algorithm is capable of delivering high-quality hyperspectral demosaicking in real-time.

\subsection{User study}
\begin{table}
\begin{center}
\begin{tabular}{|cc|cc|cc||cc|}
\hline
Linear & SGC & Linear & Ours & SGC & Ours & no RGB model & RGB model \\
\hline
26 & 135 & 17 & 144 & 30 & 131 & 52 & 155 \\
\hline
\end{tabular}
\end{center}
\caption{Number of votes received for each demosaicking method in all pairwise comparisons in the image quality assessment survey.}
\label{tab:user-study}
\end{table}
A qualitative user study was conducted to evaluate the results of our proposed demosaicking algorithm, aiming to achieve two main objectives: 
\begin{description}
    \item[Objective 1] Assess whether our algorithm produces results with improved spatial image quality compared to existing methods.
    \item[Objective 2] Determine whether our RGB model can convert hyperspectral images into RGB images with better colour, thereby aiding surgeons in making informed decisions.
\end{description}

We implemented a forced-choice pairwise comparison design \citep{mantiuk2012image}.
To avoid burdening the users with too many comparisons, based on our quantitative results, we discarded the PPID and GRMR models as they already proved inferior but otherwise used the same models as in the quantitative comparison. 

The user study comprised 30 survey questions, each presenting two images of the same scene for direct comparison. Participants were required to select one image before proceeding to the next question. These questions alternated between testing Objective 1 (comparing our method against SGC or linear, and SGC against linear -- all with a fixed sRGB conversion) and Objective 2 (comparing results of our proposed approach with either a fixed sRGB conversion or our trained RGB model).
Question ordering were randomised to prevent bias.
Each participant received exactly 7 questions for each pair of comparisons under Objective 1, and 9 questions for Objective 2, ensuring a balanced and controlled exposure to all test conditions. 

Twenty-three neurosurgical experts completed the survey. Participants included 7 consultants and 16 specialist neurosurgery trainees with 1 to 16 years of experience. The results from our user study are detailed in Table~\ref{tab:user-study}.
We applied the Bradley-Terry model \citep{bradley1952rank} to quantify the preference scales to show that our method is significantly more favored compared to SGC and linear demosaicking.
The estimated preference scales from the Bradley-Terry analysis are \( \pi = (0.053, 0.213, 0.734) \) for Linear, SGC, and Our method respectively.
The analysis revealed that our proposed algorithm is approximately 13.8 times more likely to be preferred over linear demosaicking, and about 3.5 times over SGC, with p-values indicating a statistically significant preference (\( p < 0.01 \)) for our algorithm.
Using the same method, we analysed the effectiveness of our RGB model in enhancing the colour accuracy and overall visual quality of the images.
The estimated preference scales are \( \pi = (0.251, 0.749) \), with p-value below 0.01, once again showing that participants in this study significantly prefer images from the proposed RGB model.

\section{Conclusion}
In this work, we introduced a novel GAN-based approach for medical hyperspectral image demosaicking. 
Our proposed algorithm uniquely circumvents the need for high-resolution medical hyperspectral data, which are challenging to obtain in a surgical environment. 
Instead, it uses readily available snapshot mosaic hyperspectral images and surgical microscopy RGB images. 
The performance of our hyperspectral demosaicking model has been significantly enhanced by our introduction of the IPS loss and the proposed cycle-consistent adversarial training based on RGB reconstructions. 
Additionally, the RGB visualisation has been refined through a simple MLP model, effectively compensating for the limited spectral resolution and range typically associated with snapshot mosaic cameras.
This work has undergone both quantitative and qualitative evaluation, demonstrating substantial improvements over existing self-supervised demosaicking methods, and thus proving its potential for seamless integration to real-time intra-operative surgical applications.
The feedback obtained from neurosurgical experts shows the potential practical implications of these improvements, suggesting a strong alignment with the needs of end-users in clinical environments.

\section*{Declarations}
\begin{itemize}
    \item Funding: 
    This study/project is funded by the NIHR [NIHR202114]. The views expressed are those of the author(s) and not necessarily those of the NIHR or the Department of Health and Social Care.
    This work was supported by core funding from the Wellcome/EPSRC [WT203148/Z/16/Z; NS/A000049/1].
    This project has received funding from the European Union's Horizon 2020 research and innovation programme under grant agreement No 101016985 (FAROS project).
    PL is funded by China Scholarship Council.
    For the purpose of open access, the authors have applied a CC BY public copyright license to any Author Accepted Manuscript version arising from this submission.
    \item Conflict of interest:
    TV and JS are co-founders and shareholders of Hypervision Surgical.
    \item Ethics approval:
    All procedures within this study involving human subjects were in accordance with both the institutional and regional ethical committee (REC reference 22/LO/0046,  ClinicalTrials.gov ID NCT05294185) and with the 1964 Helsinki declaration and its later amendments. 
    \item Informed consent:
    Informed consent was obtained from all individual participants involved in the study.
\end{itemize}




\section*{Supplementary material}

# Supplementary material: A self-supervised and adversarial approach to hyperspectral demosaicking and RGB reconstruction in surgical imaging

Peichao Li[1]
peichao.li@kcl.ac.uk

Oscar MacCormac[1,2]
oscar.j.maccormac@kcl.ac.uk

Jonathan Shapey[1,2]
jonathan.shapey@kcl.ac.uk

Tom Vercauteren[1]
tom.vercauteren@kcl.ac.uk

[1] School of Biomedical Engineering & Imaging Sciences
King's College London
London, UK

[2] Department of Neurosurgery
King's College Hospital NHS Foundation Trust,
London, UK

## 1 Additional Examples from the Results

Figure 1 shows some more examples to compare the RGB visualisations of different demosaicking algorithms, including linear demosaicking, PPID [3], GRMR [4], SGC [1] and our proposed GAN-based algorithm. We have also included three video examples to compare linear demosaicking and our proposed algorithm in the supplementary materials.

## 2 Per-band pixel differences plot

A plot illustrating the per-band pixel differences for one of the test images is presented in Figure 2. This plot combine with the box plot in the main paper reveal that larger differences tend to occur only in regions where our algorithm enhances spatial details, thereby confirming that these differences are not indicative of any spectral shift. Note that in this plot, gamma has been adjusted for better visualisation of small pixel value differences.

## 3 Ablation Study

Table 1 presents the results of an ablation study measuring the quality of the RGB visualisation of the demosaicked hyperspectral images. Detailed ablation study of the SR losses has already been covered in the supplementary material of our previous paper [1], so we won't repeat it and will focus solely on the effects of IPS loss and adversarial losses only. We evaluated our proposed demosaicking algorithm by selectively incorporating either the

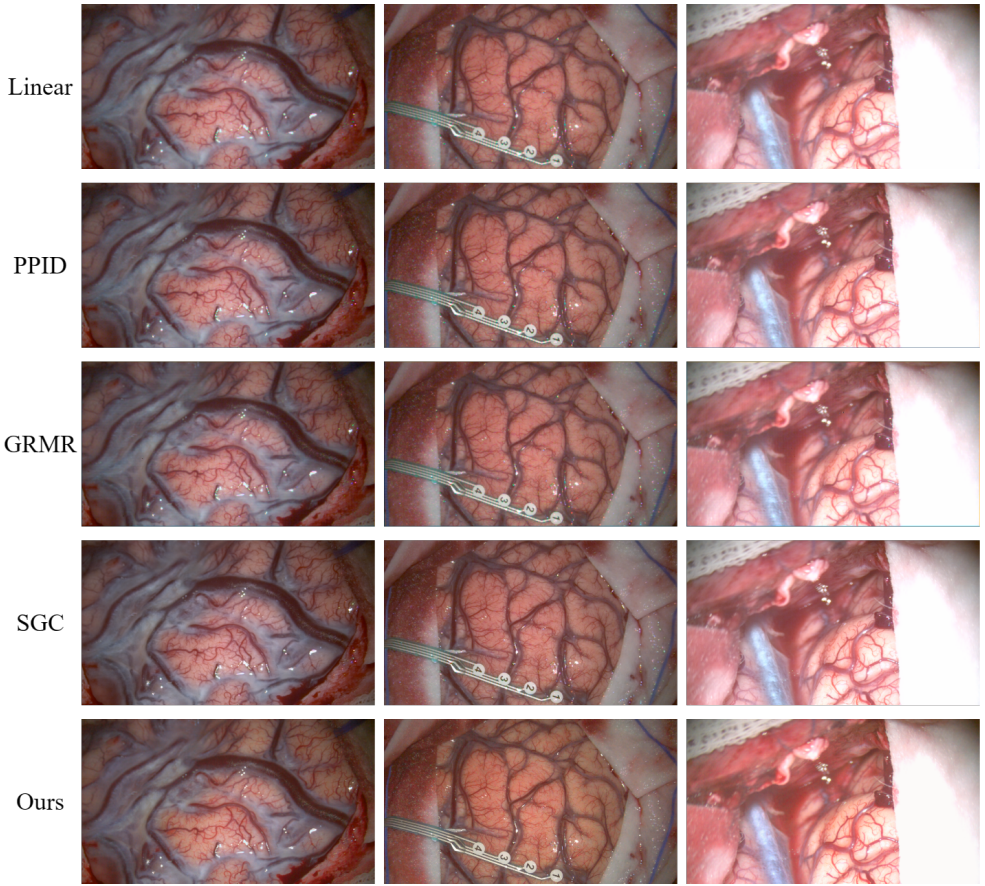

Figure 1: Comparison between different demosaicking methods on example NeuroHSI test images.

IPS loss, the adversarial losses (PatchGAN loss and cycle consistency loss), or both. In all experiments, we retained other loss terms, including total variation and SGC terms, by default. In the experiment with only the IPS loss, the weighting terms were adjusted to $\lambda_{\text{TV}} = 1 \times 10^{-3}$, $\lambda_{\text{IPS}} = 1$, and $\lambda_{\text{SGC}} = 1 \times 10^{-2}$. In the experiment with only the adversarial losses, the weighting terms remained unchanged. An illustrative example of these different experiments can be found in Figure 3. It is evident that both IPS loss and adversarial losses are crucial components of our proposed demosaicking algorithm. When only the IPS loss was used, the weighting for the SGC had to be reduced to avoid artefacts caused by the SGC term over-promoting spatial correlations, which inevitably diminished the ability to recover image sharpness. When only the adversarial losses were applied, the example image clearly shows that although the results exhibit good sharpness, noticeable gridding artefacts appear, which can be mitigated by the IPS loss.

We also compared the adversarial training with fixed parameters of the RGB models to the training where the RGB models were also trained. Although training with the RGB model achieved a better BRISQUE score, as shown in Table 1, the p-value of 0.21 indicates that this difference is not statistically significant. This result is expected, as BRISQUE primarily

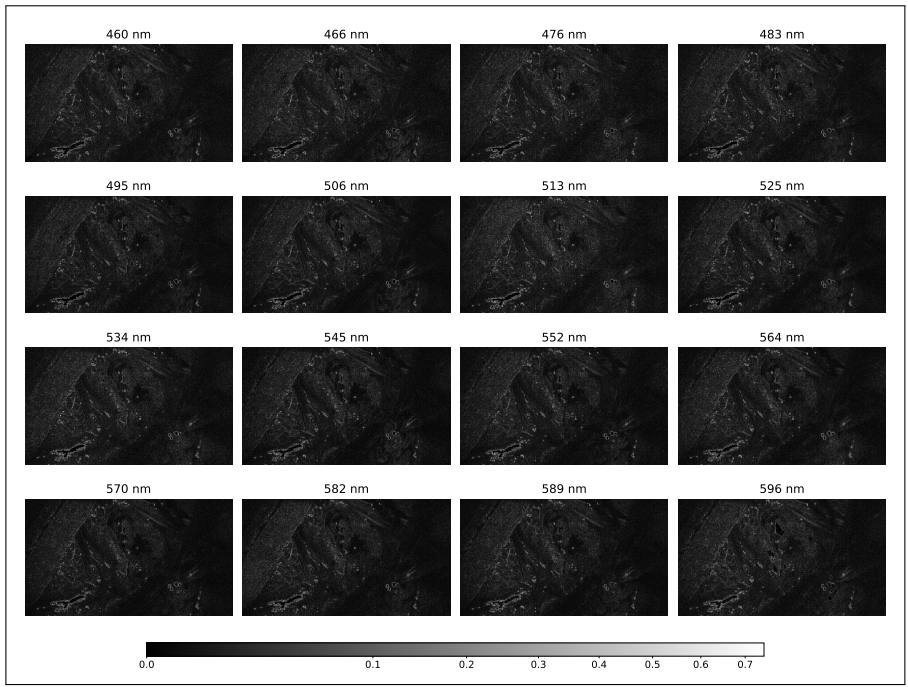

Figure 2: Illustration of per-band pixel difference between our demosaicking results and linear demosaicking on one of the test images.

focuses on assessing the spatial naturalness of images. Our proposed colour model only involves $1 \times 1$ convolutions, which do not extract spatial features and thus do not have major contribution to the spatial quality of the images. The main contribution of our proposed RGB model is to improve the colour fidelity of the RGB visualisation of the demosaicked image, which has been evaluated through our user study described in the main paper.

## 4 More information on the survey

We developed a web application for our user survey, as shown in Figure 4. According to [2], forced-choice pairwise comparison is the fastest and most accurate method for image quality assessment, so we designed a two-alternative forced-choice (2AFC) image quality survey,

| Method | BRISQUE | FID Score |
|---|---|---|
| IPS only | $62.26 \pm 6.18$ | 98.02 |
| GAN only (RGB model) | $33.04 \pm 8.26$ | 86.20 |
| IPS + GAN (fixed RGB model) | $19.70 \pm 5.91$ | 79.02 |
| IPS + GAN (trained RGB model) | $19.35 \pm 5.77$ | 75.62 |

Table 1: Ablation study results measuring the quality of the RGB visualization of the demosaicked hyperspectral images. Lower is better for both BRISQUE and FID.

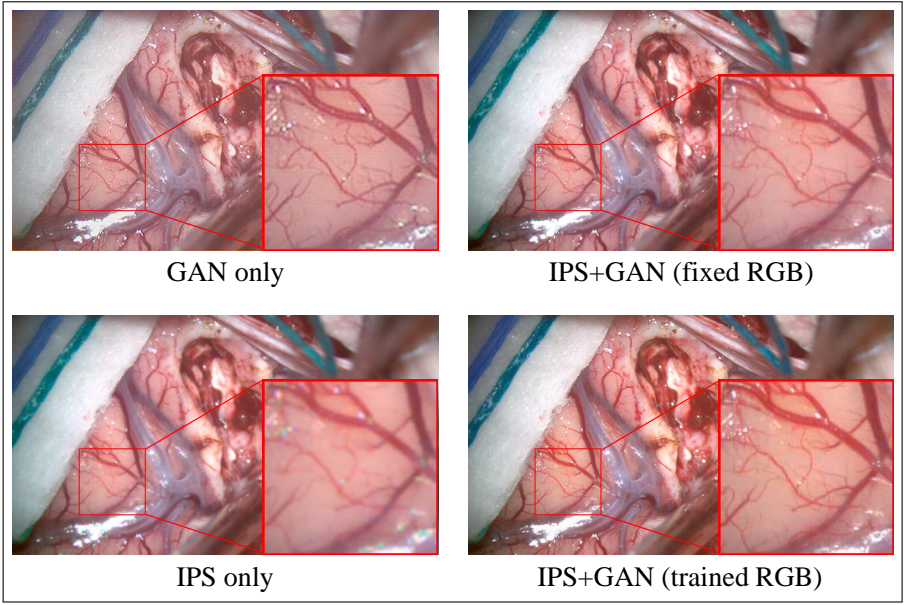

Figure 3: An example from the test set showing the effects of different loss terms in the ablation study.

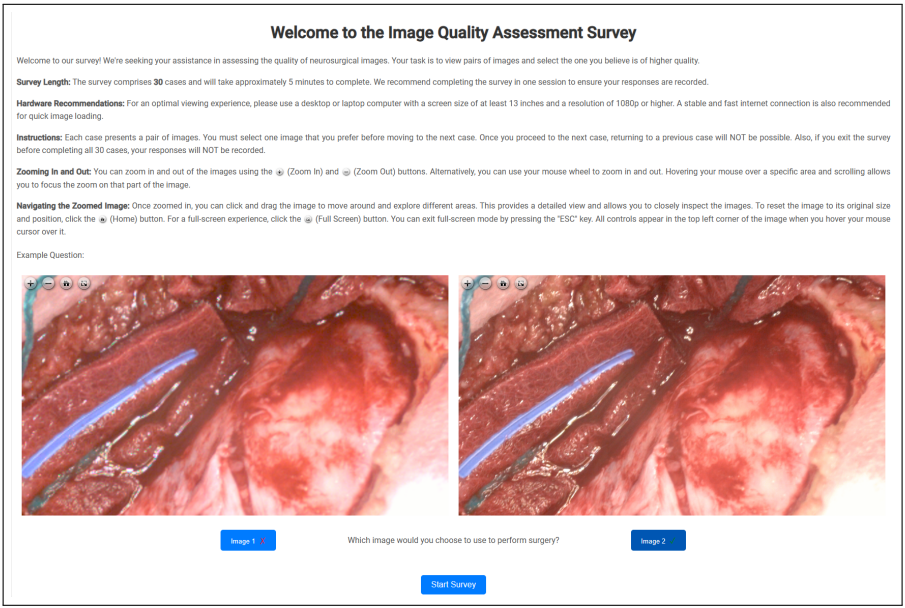

Figure 4: The instruction page of our survey app with an example question. Zoom option is provided to help the participants with observing the images in more details.

where observers were required to compare two images at a time and select the one with better quality without using any rating scales. In this web app, participants were presented with two images per question, allowing them to take their time and zoom in to examine details. They were asked to choose one image with better quality, and once a choice was made, they could not go back and alter their answers. Upon completion of the survey, the data were stored in the backend of the web app for analysis. Using this web app, we collected responses from 23 neurosurgical experts, and the results are presented in the main paper.



\end{document}